# Nickelate superconductivity without rare-earth magnetism: (La,Sr)NiO$_2$

Motoki Osada,[1,2]* Bai Yang Wang,[1,3] Berit H. Goodge,[4,5] Shannon P. Harvey,[1,2] Kyuho Lee,[1,3] Danfeng Li,[1,2,6] Lena F. Kourkoutis,[4,5] and Harold Y. Hwang[1,2]*

[1]*Stanford Institute for Materials and Energy Sciences, SLAC National Accelerator Laboratory, Menlo Park, CA 94025, United States.*

[2]*Department of Applied Physics, Stanford University, Stanford, CA 94305, United States.*

[3]*Department of Physics, Stanford University, Stanford, CA 94305, United States.*

[4]*School of Applied and Engineering Physics, Cornell University, Ithaca, NY 14853, United States.*

[5]*Kavli Institute at Cornell for Nanoscale Science, Cornell University, Ithaca, NY 14853, United States.*

[6]*Department of Physics, City University of Hong Kong, Kowloon, Hong Kong, People's Republic of China.*

*Correspondence to: mosada@stanford.edu, hyhwang@stanford.edu

**ABSTRACT**

The observation of superconductivity in infinite layer nickelate (Nd,Sr)NiO$_2$ thin films has led to rapid theoretical and experimental investigations of these copper-oxide-analogue systems [1-15]. Superconductivity has also been found in (Pr,Sr)NiO$_2$ [16,17], but not previously in (La,Sr)NiO$_2$ [2], raising a fundamental question whether superconductivity is associated with the presence of rare-earth moments [18,19]. Here we show that with significant materials optimization, substantial portions of the La$_{1-x}$Sr$_x$NiO$_2$ phase diagram can enter the regime of coherent low-temperature transport ($x$ = 0.14 – 0.20), with subsequent superconducting transitions and a maximum onset of ~ 9 K at $x$ = 0.20. Additionally, we observe the unexpected indication of a superconducting ground state in undoped LaNiO$_2$, which likely reflects the self-doped nature of the electronic structure. Combining the results of (La/Pr/Nd)$_{1-x}$Sr$_x$NiO$_2$ reveals a generalized superconducting dome, characterized by systematic shifts in the unit cell volume and in the relative electron-hole populations across the lanthanides.

**MAIN TEXT**

Across the lanthanide series, systematic studies of the infinite layer nickelates using density-functional theory (DFT)-based approaches have generally shown a smooth evolution of the electronic structure, with little qualitative dissimilarity [20,21]. While these approaches treat the 4$f$ electrons as part of the lanthanide core, the observation of superconductivity in (Nd,Sr)NiO$_2$, but not in (La,Sr)NiO$_2$, has led to consideration of 4$f$-orbital hybridization to account for this distinction [18,19]. In parallel with these theoretical efforts, significant advances have been made in the understanding and optimization of thin film synthesis of the Nd/Pr based systems [11,16]. Important aspects include improving the crystallinity of the strained perovskite precursor phase, as well as the subsequent oxygen deintercalation reaction. As a result, the magnitude of the normal state resistivity for optimized samples is significantly lower than in several prior reports [22-25], consistent with these advances in materials quality. Furthermore, a key observation is that the low temperature normal state resistivity in Nd/Pr based infinite layer nickelate films must be below the Mott-Ioffe-Regel limit to exhibit a superconducting transition [12,17]. Given the conceptual importance of the presence or absence of superconductivity in (La,Sr)NiO$_2$, and recent materials improvements, we revisit the synthesis of La$_{1-x}$Sr$_x$NiO$_2$ thin films epitaxially stabilized on SrTiO$_3$ substrates and the study of its phase diagram.

Figure 1a shows the temperature-dependent resistivity $\rho(T)$ for La$_{1-x}$Sr$_x$NiO$_2$ thin films ($x$ = 0, 0.1, 0.2, and 0.3) from the initial study [2]. In all cases, the samples exhibited weak semiconducting behavior, with $x$ = 0.2 showing the highest conductivity overall. The notion of increasing the electronic bandwidth, to further increase conductivity, motivated the subsequent investigation of (Nd,Sr)NiO$_2$ with the smaller cation Nd$^{3+}$ in place of La$^{3+}$. The undoped LaNiO$_2$ sample can be compared with other literature reports [22-25], spanning almost two orders of magnitude in resistivity variation. While the resistivities of the initial samples were comparable to several of the prior reports [22,23], others showed lower resistivity [24,25]. This indicated the possibility that the resistivities observed, and by extension for the doped samples, might not be intrinsic, but limited by materials imperfections. To explore this possibility, we worked to improve the crystallinity of (La,Sr)NiO$_2$ films following the approach recently taken for the Nd system [11] (Extended Data Figs. 1 and 2), which led to pulsed laser deposition growth conditions at higher oxygen pressure, higher laser fluence, and smaller laser spot size (see Methods for details). For comparison, the optimized $x = 0.2$ sample is also shown in Fig. 1a, which shows substantially lower resistivity and metallic temperature dependence, followed by a superconducting transition at low temperatures.

This significant change in transport properties can be associated with changes in the crystallinity. In Fig. 1b, we show x-ray diffraction (XRD) $\theta$-$2\theta$ wide scans for initial (non-superconducting; top) and optimized (superconducting; bottom) $x = 0.2$ samples, in which the two film peaks corresponding to (001)-oriented La$_{0.8}$Sr$_{0.2}$NiO$_2$ can be seen. Firstly, we observe a much more prominent 001 film peak, which has been found to be an important proxy for the crystallinity of the infinite layer phase [11]. Furthermore, as shown in Fig. 1c, the full-width-at-half-maximum (FWHM) of rocking curve measurements on the 002 peaks of La$_{0.8}$Sr$_{0.2}$NiO$_2$ are dramatically narrowed, from broad (~ 0.9 deg, initial) to sharp (~ 0.1 deg, optimized). Figure 1d shows a high-angle annular dark-field (HAADF) scanning transmission electron microscopy (STEM) cross-sectional image for the optimized $x = 0.2$ sample. Overall, a coherent infinite layer structure is observed, with Ruddlesden-Popper (RP)-type structural faults, rather similar to those observed in the Nd/Pr systems [11,16]. Elemental electron energy-loss spectroscopic (EELS) maps taken at the La-$M_{4,5}$, Sr-$M_{2,3}$, and Ti-$L_{2,3}$ edges indicate atomically uniform films with no chemical segregation (Fig. 1e).

Such materials improvements can be systematically extended to a wide range of compositions La$_{1-x}$Sr$_x$NiO$_2$ ($0 \leq x \leq 0.30$). The formation of high-quality infinite layer structure across all $x$ values studied here has been confirmed by XRD $\theta$-$2\theta$ scans (Extended Data Fig. 1a and c). Figure 2 displays the temperature dependent resistivity measured down to dilution refrigerator temperatures. At low and high doping levels ($x = 0.04, 0.06, 0.10, 0.24$, and $0.30$), the resistivity curves show high-temperature metallic behavior, with an approximately logarithmic upturn at low temperatures [26], analogous to that in the Nd/Pr systems [12,13,17]. Note that $x = 0.24$ and $0.30$ show some evidence for decreasing crystallinity in wider peaks in rocking curve measurements (Extended Data Fig. 1b and d), likely reflecting the increasingly unstable high oxidation state of the perovskite precursor phase. Superconductivity emerges at intermediate doping ($0.14 \leq x \leq 0.20$), with a maximum superconducting transition temperature $T_c$ onset (defined as 90% of the 20 K resistivity, $T_{c,90\%R}$) of 9 K at $x = 0.20$. Thus, we find that superconductivity in infinite layer nickelates is not dependent on rare-earth states, namely, $4f$ electrons and their corresponding (local) magnetic moments. Rather, the key observation is that with increased crystallinity, the resistivity is reduced, such that substantial portions of the (La,Sr)NiO$_2$ phase diagram are in the regime of coherent low-temperature transport, as indicated by the scale of the quantum sheet resistance $h/e^2$ per NiO$_2$ plane (black dots in Fig. 2e).

Not only does the improved crystallinity reveal a superconducting dome, perhaps even more surprisingly, the optimized undoped LaNiO$_2$ is also in the coherent transport regime (Fig. 2e), and shows an incomplete superconducting transition down to lowest measurement temperatures. These results suggest that at sufficiently low disorder, the ultimate ground state of undoped infinite nickelates may be a superconducting state. This intriguing circumstance likely reflects the self-doped nature of the electronic structure, involving both partially occupied electron pockets primarily derived from the rare-earth $5d$ states, and the two-dimensional hole band of mainly Ni $3d_{x2-y2}$ character, which is distinct from the copper oxides [1,3,5,9,26,27]. Furthermore, we note a qualitative similarity to the expansion of superconducting phases in twisted bilayer graphene with decreasing disorder [28,29]. At lowest measured temperatures, all infinite layer nickelate samples (optimized or not), and all doping levels, are either superconducting or weakly insulating.

In addition to the surprisingly low resistivity scale of these optimized $LaNiO_2$ samples, another qualitative distinction to prior literature (and our initial growth conditions) is that we do not observe a $c$-axis to $a$-axis reorientation transition with continued reduction [23,30]. For instance, an optimized $LaNiO_2$ thin film (7 nm thick) is robust up to 300 ºC, and directly decomposed from the $c$-axis structure after 22 hours of reduction (Extended Data Fig. 2). This is in contrast to the $a$-axis transition that was observed after 3 hours of reduction at 280 ºC in a prior report [30]. To understand both the structure and low resistivity, Fig. 3a shows the cross-sectional STEM image of optimized $LaNiO_2$. First we see that for large regions the film displays excellent crystallinity, largely free of the RP-type faults and other extended defects commonly observed in $(Nd,Sr)NiO_2$, $(Pr,Sr)NiO_2$, prior studies of $LaNiO_2$, and the optimized $x = 0.20$ shown in Fig. 1d [11,16,25]. This partially reflects the closer lattice match between the precursor $LaNiO_3$ and $SrTiO_3$ substrate, and the improvements in growth. The second general feature is the presence of diagonal extended defects (Fig. 3).

Local mapping indicates that this effective domain boundary is consistent with a partial rotation from $c$-axis to $a$-axis and back to $c$-axis. Using real space wave-fitting analysis [31], we track the rotation of the out-of-plane lattice vector and the local in-plane and out-of-plane lattice constants. At the diagonal extended defect, the out-of-plane lattice vector rotates by 5-10° relative to the average orientation across the field of view. Similar rotations are also observed in the in-plane lattice vectors (Extended Data Fig. 4), suggesting the two rotate together (rather than an independent rotation which could arise from large lattice shear). This rotation is accompanied by a concurrent change in the ratio of in-plane and out-of-plane lattice constants, which indicates a local reorientation of the $c$ and $a$ axes. Throughout most of the film, the observed ratio between in-plane and out-of-plane lattice constants is on the order of 1.1-1.2, approximately the nominal ratio of the bulk $LaNiO_2$ $a$ (3.96Å) and $c$ (3.38 Å) axis values [32]. At the diagonal defects, however, the in-plane versus out-of-plane ratio is approximately 0.8-0.9, comparable with the inverse ratio ($c/a$), suggesting that these regions are locally reorienting towards an out-of-plane $a$-axis. Given that the $a$-axis lattice constant of $LaNiO_2$ is slightly larger than that of $SrTiO_3$ (3.91 Å), while the $c$-axis lattice constant is smaller than that of $SrTiO_3$, we interpret this local rotation as a compensation mechanism for the in-plane strain. Based on these observations, we infer that the tendency towards a reorientation transition remains, but it is suppressed with increasing crystallinity. This is presumably the remaining extended defect to be controlled for further improvements in materials quality, towards even cleaner $LaNiO_2$.

Finally, we discuss these optimized (La,Sr)NiO$_2$ results in the broader context of the doped infinite layer nickelate films studied thus far. Together with the observation of superconductivity here, we can plot all existing experimental phase diagrams in Fig. 4a. We note that while a small dip of $T_{c,90\%R}$ is observed in the superconducting domes of $La_{1-x}Sr_xNiO_2$ and $Nd_{1-x}Sr_xNiO_2$, it is absent in $Pr_{1-x}Sr_xNiO_2$ [12,13,17]. These results could suggest that a small distinction in the strain state might affect the ground state of the nickelates; for example, phases such as stripe order, widely observed in copper oxides [33,34] and subject to electronic correlations and the local chemical environment [35], might be similarly present. Conversely, the strong disorder dependence of superconductivity observed here suggests that the detailed structure of the superconducting dome may yet reflect materials imperfections.

The variation in chemical composition impacts the interplanar spacing, which can be used to consolidate the different phase diagrams. In Fig. 4b, we plot $T_{c,90\%R}$ with respect to the measured $c$-axis lattice constant (in all cases, the in-plane lattice is clamped to the $SrTiO_3$ substrate). This is akin to assuming that only the unit cell dimensions are relevant for the variations observed. The resultant plot is highly suggestive of the existence of a generalized dome of superconductivity, across which different specific compounds cross. Note that this consideration neglects systematic trends in the electronic structure arising from Ni $3d$ – rare-earth $5d$ hybridization across the lanthanide series found in electronic structure calculations [20,21]. To examine this point, we have performed normal state Hall effect measurements for $La_{1-x}Sr_xNiO_2$ ($0 \leq x \leq 0.30$) (Fig. 4c). Similar to Pr/Nd systems, the normal state Hall coefficient $R_H$ of $La_{1-x}Sr_xNiO_2$ increases with Sr substitution. Above intermediate doping levels ($x \geq 0.14$), $R_H(T)$ monotonically increases with decreasing temperature. In particular, for samples with higher doping ($x \geq 0.24$), $R_H(T)$ crosses zero in a consecutive manner.

Comparatively, in Fig. 4d, we show $R_H(T)$ as a contour map across the available lanthanide series. In a simple two-band picture, this corresponds to a systematic shift in the relative compensation point between electron-hole populations (scaled by their mobilities) going from Nd to Pr and to La. These trends, however, are different from DFT-based calculations across the lanthanides that suggest the electron pockets grow from La to Pr/Nd [21]. One competing aspect is the effect of compressive strain, which tends to increase the electron pocket occupation and may dominate the evolution of the electronic structure [20]. $La_{1-x}Sr_xNiO_2$ is indeed more compressively strained on $SrTiO_3$ substrates than $Pr_{1-x}Sr_xNiO_2$ and $Nd_{1-x}Sr_xNiO_2$. Further investigation of these competing tendencies may lead to an understanding of the underlying structure of the superconducting dome (Fig. 4b), and the absence of superconductivity in bulk samples [36,37].

**NOTE**

During the preparation of this manuscript, we became aware of a report of superconductivity in $(La,Ca)NiO_2$ [38].

**METHODS**

**Thin film growth.** Solid-solution $La_{1-x}Sr_xNiO_3$ nickelate thin films (5.5 – 10 nm thick) were grown on single-crystalline $SrTiO_3$ (001) substrates by pulsed laser deposition using stoichiometric polycrystalline targets ablated with a KrF excimer laser (wavelength = 248 nm), followed in some cases by the growth of ~2 nm $SrTiO_3$ capping layer. Prior to the growth, $SrTiO_3$ substrates were pre-annealed at 930 °C under an oxygen partial pressure of $5 \times 10^{-6}$ Torr, in order to obtain an atomically flat surface topography. During growth the substrate temperature was kept at 570 °C. In our initial growth conditions [2], $La_{1-x}Sr_xNiO_3$ thin films were grown under oxygen pressure of 35 mTorr using the laser fluence of 1.47 J/cm² (laser spot area 3.9 mm²); after optimization, the growth conditions were a laser fluence of 1.39 J/cm² and 2.19 J/cm² for undoped $LaNiO_3$ and doped $La_{1-x}Sr_xNiO_3$ ($x \neq 0$), respectively (laser spot area 2.6 mm²). The oxygen pressure during the growth was 200 mTorr and 250 mTorr for undoped LaNiO$_3$ and doped La$_{1-x}$Sr$_x$NiO$_3$ ($x \neq 0$). For SrTiO$_3$ capping layer growth, a laser fluence of 0.87 J/cm² and oxygen pressure of 200 mTorr were used. The laser repetition was 4 Hz for all cases.

**Topotactic reduction.** In order to obtain films of the infinite layer nickelate La$_{1-x}$Sr$_x$NiO$_2$, the as-grown precursor perovskite thin films were loosely wrapped with aluminum foil and placed in Pyrex glass tubes with calcium hydride (CaH$_2$) powder (~ 0.1 g). The glass tubes were evacuated to below 0.1 mTorr using a rotary pump and subsequently sealed using a hydrogen torch. The sealed tube was then heated in a tube furnace at 260 °C for 60 min for SrTiO$_3$ capped films and at 240 °C for 60 min for uncapped films unless otherwise specified (e.g. the reduction study of Extended Data Fig. 2). The thermal ramping and cooling rates were 10 °C/min.

**Thin film characterization.** The x-ray diffraction (XRD) $\theta$-$2\theta$ profiles and rocking curves were collected using monochromatic Cu $K_{\alpha 1}$ radiation source. The rocking curve measurements were performed for the 002-diffraction peaks of La$_{1-x}$Sr$_x$NiO$_2$. To compare the sample quality, the intensities were normalized. The longitudinal and Hall resistivity of the 2.5 mm × 5 mm films were measured using a standard six-point geometry with Al wire bonded contacts. Normal state Hall coefficients were obtained from the Hall resistivity $\rho_{yx}$ that showed linear magnetic field dependence up to 9 T.

**Scanning transmission electron microscopy and analysis.** Cross-sectional scanning transmission electron microscopy (STEM) specimens were prepared using a standard focused ion beam (FIB) lift-out process on a Thermo Scientific Helios G4 UX FIB. Samples were imaged on an aberration-corrected FEI Titan Themis microscope operated at 300 keV with a 30 mrad probe-forming convergence angle and a 68 mrad inner collection angle. Electron energy-loss spectroscopy (EELS) was performed on the same Titan Themis system equipped with a 965 GIF Quantum ER and a Gatan K2 Summit direct electron detector operated in electron counting mode. Analysis of local lattice orientation and spacing was performed using the local wave-fitting method described in [31] with a Fourier mask size correlating to 2 nm real-space resolution.

## ACKNOWLEDGMENTS

We thank S. Raghu for discussions. This work was supported by the US Department of Energy, Office of Basic Energy Sciences, Division of Materials Sciences and Engineering, under contract number DE-AC02-76SF00515, and the Gordon and Betty Moore Foundation's Emergent Phenomena in Quantum Systems Initiative through grant number GBMF9072 (synthesis equipment and dilution refrigerator measurements). B.H.G. and L.F.K. acknowledge support by the Department of Defense Air Force Office of Scientific Research (no. FA 9550-16-1-0305). This work made use of the Cornell Center for Materials Research (CCMR) Shared Facilities, which are supported through the NSF MRSEC Program (no. DMR-1719875). The FEI Titan Themis 300 was acquired through no. NSF-MRI-1429155, with additional support from Cornell University, the Weill Institute, and the Kavli Institute at Cornell. The Thermo Fisher Helios G4 UX FIB was acquired with support by National Science Foundation (DMR-1539918, DMR-1719875).

## AUTHOR CONTRIBUTIONS

M.O. and H.Y.H. conceived the experiment. M.O., K.L., and D.L. grew the nickelate films, and conducted the reduction experiments and structural characterization. B.H.G. and L.F.K. conducted electron microscopy and analysis. B.Y.W. and S.P.H. performed the transport measurements, and analysis with M.O., K.L., D.L., and H.Y.H. M.O., D.L., and H.Y.H. wrote the manuscript with input from all authors.

**COMPETING INTERESTS**

The authors declare no competing interests.

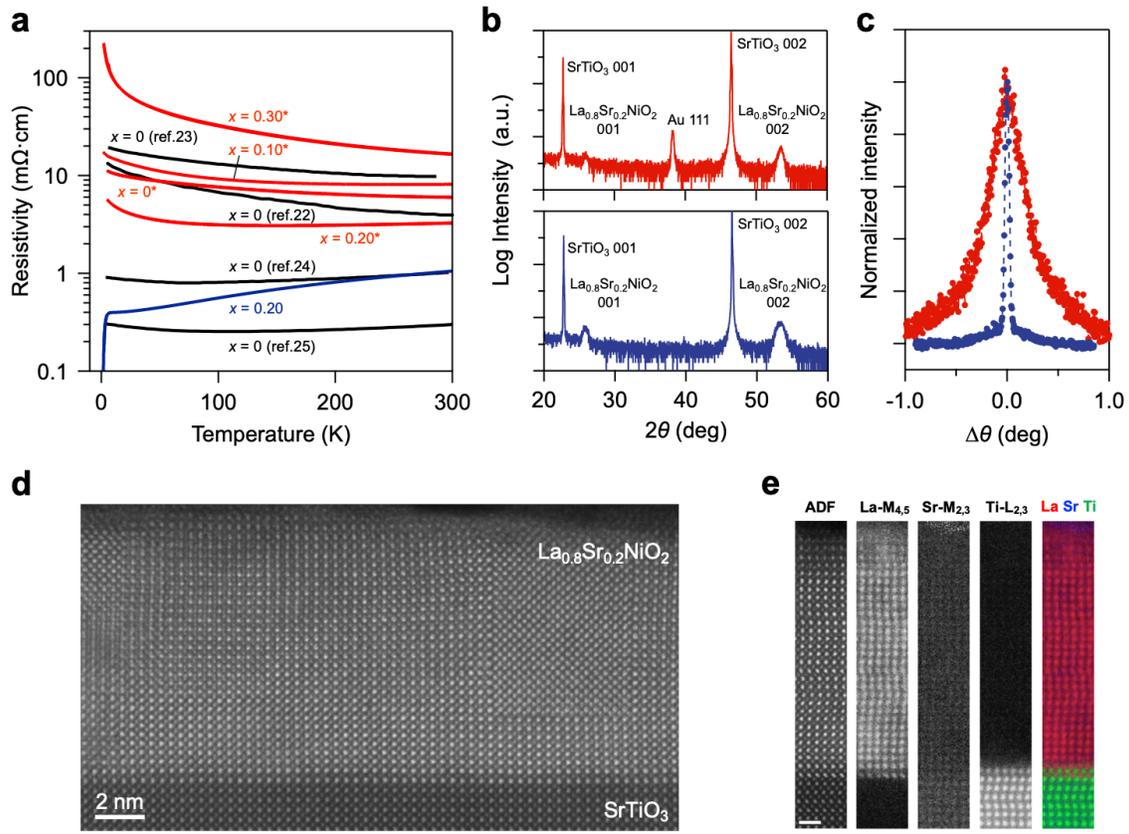

**Fig. 1 | Characteristics of initial and optimized infinite layer La$_{1-x}$Sr$_x$NiO$_2$ films. a**, Temperature-dependent resistivity curves. Data for the films synthesized under initial conditions ($x$ = 0, 0.1, 0.2 and 0.3) are shown in red with asterisk marks. Data for an optimized film ($x$ = 0.2) is shown in blue. For comparison, data for $x$ = 0 in previous reports are also shown in black, adapted from refs. [22], [23], [24], and [25]. **b**, X-ray diffraction $\theta$–$2\theta$ symmetric scans of La$_{0.8}$Sr$_{0.2}$NiO$_2$ on SrTiO$_3$ (001) substrates and **c**, rocking curves for the 002 peaks of La$_{0.8}$Sr$_{0.2}$NiO$_2$ films grown via initial (red curve; top) and optimized (blue curve; bottom) conditions. **d**, Cross-sectional high-angle annular dark-field (HAADF) scanning transmission electron microscopy (STEM) image of the optimized La$_{0.8}$Sr$_{0.2}$NiO$_2$ film on SrTiO$_3$ substrate. **e**, Annular dark field (ADF) STEM image, simultaneously acquired elemental maps of the La-M$_{4,5}$, Sr-M$_{2,3}$ and Ti-L$_{2,3}$ edges, and false-colored composite image for which La, Sr, and Ti are shown in red, blue, and green, respectively. Scale bar indicates 1 nm.

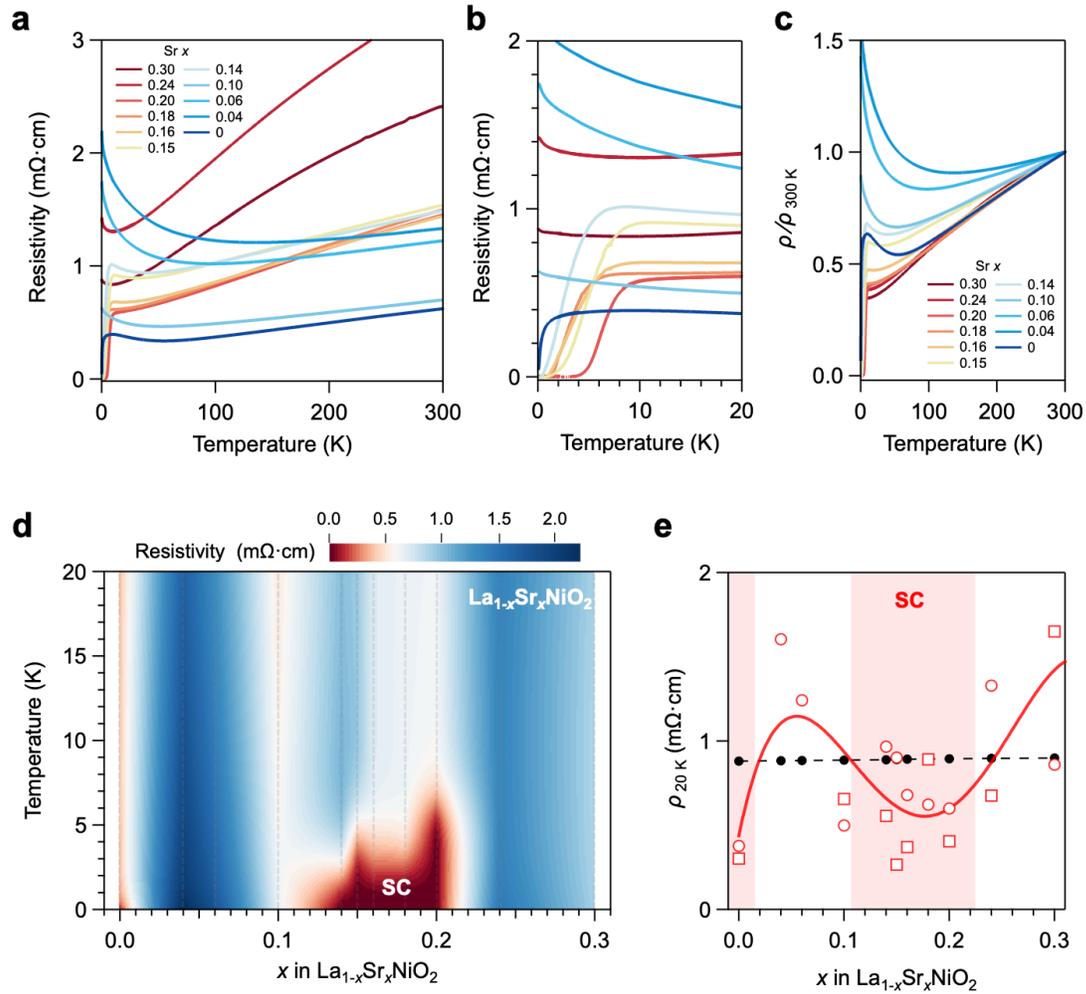

**Fig. 2 | Transport properties and phase diagram of optimized thin film La$_{1-x}$Sr$_x$NiO$_2$. a**, Temperature dependent resistivity curves for La$_{1-x}$Sr$_x$NiO$_2$ thin films upon growth optimization. **b**, The enlarged $\rho(T)$ curves below 20 K. **c**, The $\rho(T)$ data normalized by the resistivity value at 300 K ($\rho/\rho_{300K}$). **d**, The phase diagram of La$_{1-x}$Sr$_x$NiO$_2$ thin films, extracted from the $\rho(T)$ measurements. Dashed lines indicate the measured doping ratios. A superconducting (SC) dome is observed between $x = 0.14$ and 0.20, bounded by weakly insulating regimes. **e**, Resistivity as a function of Sr $x$ at 20 K, extracted from (**a**) (open circles), with additional data from uncapped samples (open squares, Extended Data Fig. 3). The red solid curve is a guide to the eye, and black dots connected with a dashed line represent resistivity values corresponding to a quantum sheet resistance per NiO$_2$ plane.

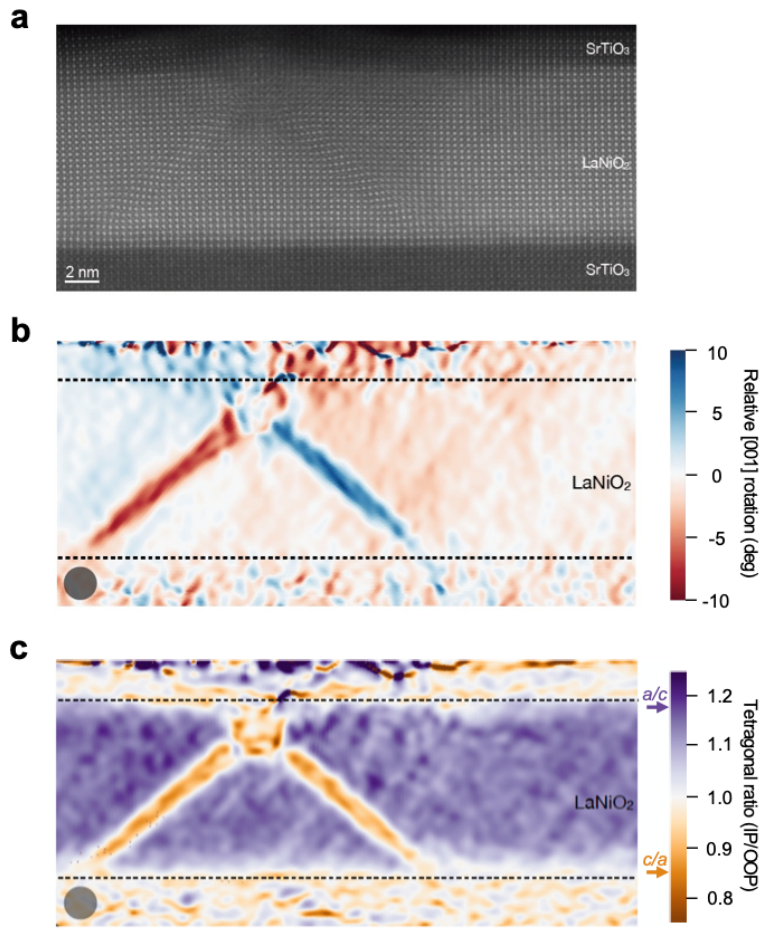

**Fig. 3 | Microscopic structure of the optimized undoped LaNiO$_2$ film. a**, Cross-sectional HAADF-STEM image of a LaNiO$_2$ film on a SrTiO$_3$ substrate with a SrTiO$_3$ capping layer. **b**, Corresponding wave-fitting out-of-plane lattice rotation map tracking [001] for SrTiO$_3$/LaNiO$_2$. Angles are measured relative to the map average. **c**, Map of the local ratio between in-plane (IP) and out-of-plane (OOP) lattice spacings. The purple and orange arrows on the color bar mark the nominal *a/c* and *c/a* ratios of bulk LaNiO$_2$, respectively. Dotted lines mark the interfaces between LaNiO$_2$ film and SrTiO$_3$ substrate/capping layer. The grey dot shows the real-space resolution of the Fourier mask used to generate the wave amplitude images used for (**b**) and (**c**).

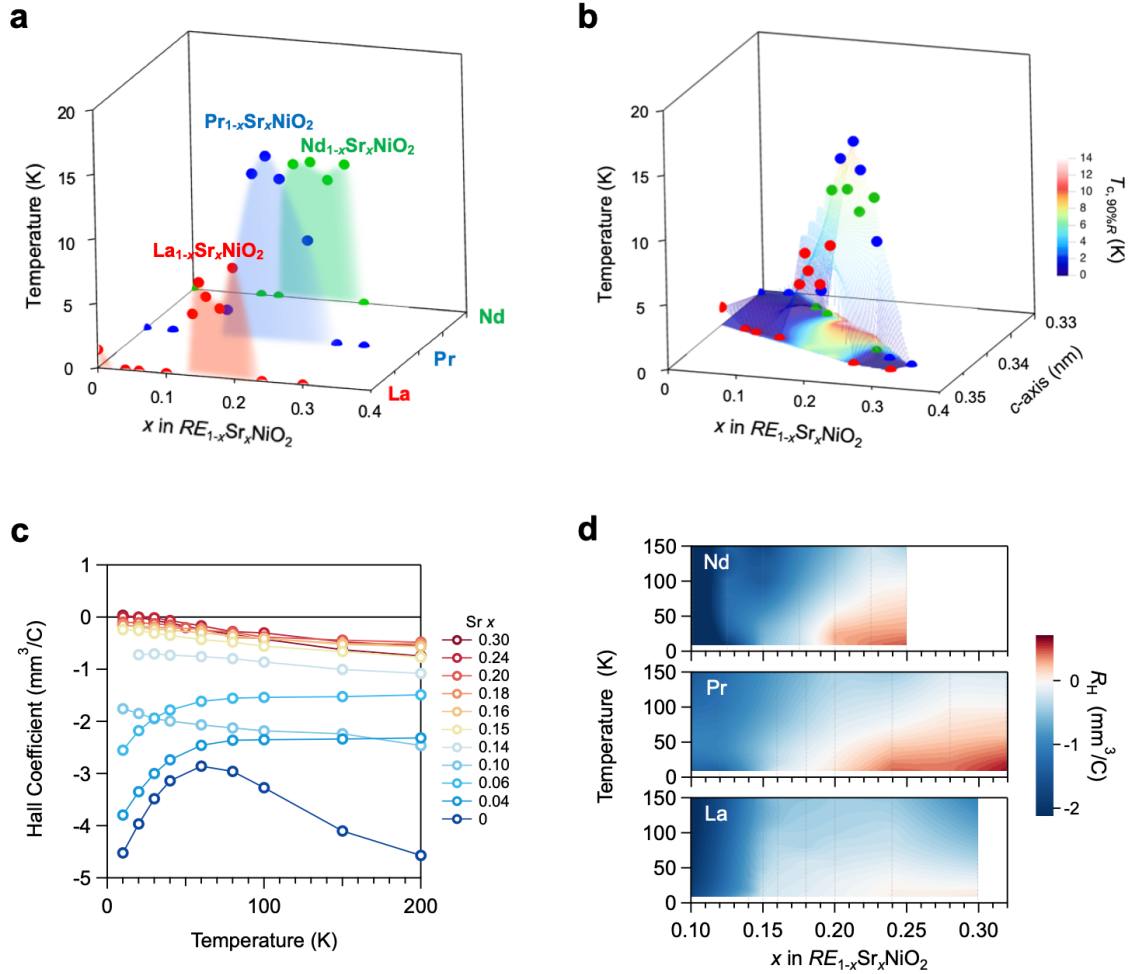

**Fig. 4 | Comparison of the superconducting phase diagram and the Hall coefficients of various doped infinite layer nickelates. a**, $A$-site cation dependent superconducting domes for nickelate thin films ($A_{1-x}Sr_xNiO_2$; $A$: La, Pr [17], Nd [12]). **b**, The superconducting domes as a function of $c$-axis lattice constant, indicative of a unified phase diagram. $A$: La (red), Pr (blue), and Nd (green). **c**, Normal-state Hall coefficient as a function of temperature for optimized $La_{1-x}Sr_xNiO_2$ thin films. **d**, Contour maps of the Hall coefficients of Nd-, Pr- and La-based infinite layer nickelates. The overlaid dashed lines indicate the measured doping levels.

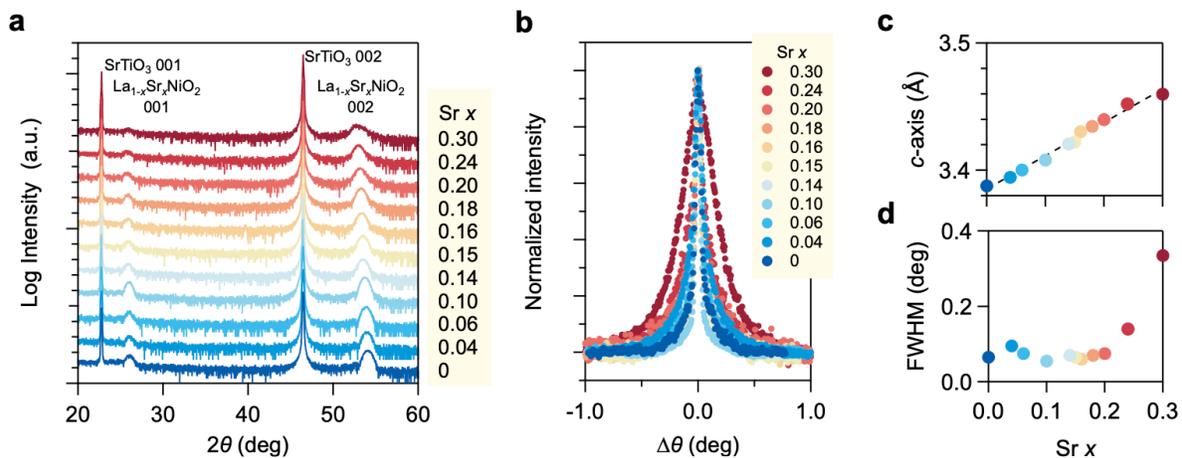

**Extended Data Fig. 1. a**, X-ray diffraction $\theta$–$2\theta$ symmetric scans of the optimized La$_{1-x}$Sr$_x$NiO$_2$ thin films on SrTiO$_3$ (001) substrates and **b**, rocking curves for the 002 film peaks. **c**, The extracted $c$-axis lattice constants of La$_{1-x}$Sr$_x$NiO$_2$ thin films and **d**, full-width-at-half-maximum (FWHM) of the rocking curves in (**b**).

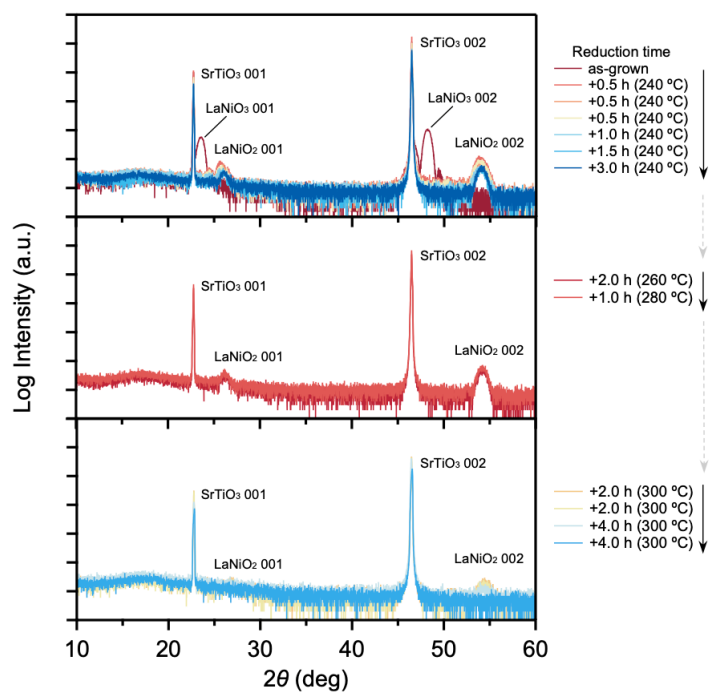

**Extended Data Fig. 2.** X-ray diffraction $\theta$–$2\theta$ symmetric scans of optimized as-grown LaNiO$_3$ on SrTiO$_3$ (001) substrate and after a series of reduction steps. The same sample was subsequently reduced at 240 - 300 °C for various times, labeled accordingly. The arrows in the legends indicate the reduction history.

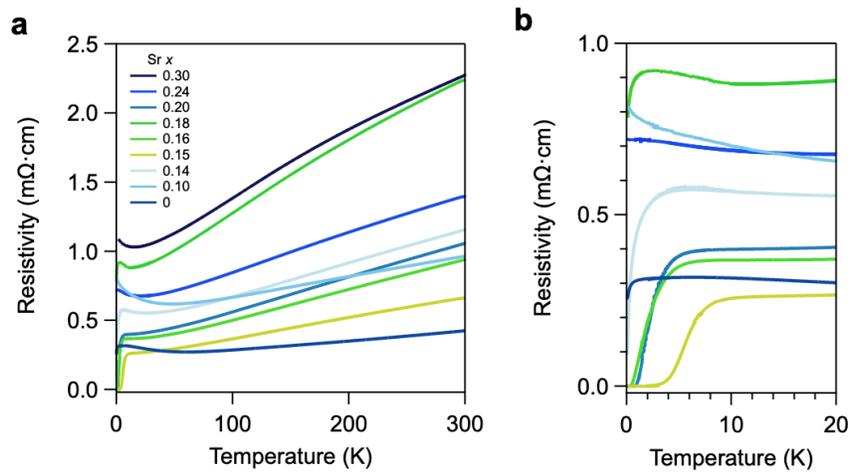

**Extended Data Fig. 3. a**, Temperature dependent resistivity curves for a set of optimized $La_{1-x}Sr_xNiO_2$ thin films without $SrTiO_3$ capping layers. **b**, The enlarged $\rho(T)$ curves below 20 K.

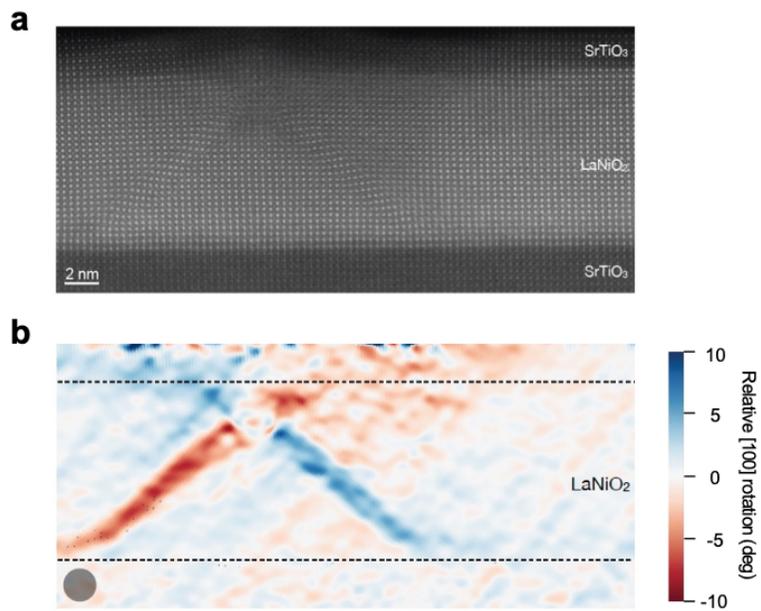

**Extended Data Fig. 4. a**, Cross-sectional HAADF-STEM image of a LaNiO$_2$ film on a SrTiO$_3$ substrate with a SrTiO$_3$ capping layer (same data as Fig. 3a), and **b**, corresponding wave-fitting in-plane lattice rotation map tracking [100] for SrTiO$_3$/LaNiO$_2$. Angles are measured relative to the map average. Dotted lines mark the interfaces between LaNiO$_2$ film and SrTiO$_3$ substrate/capping layer. The grey dot shows the real space resolution of the Fourier mask used to generate the wave amplitude images used for (**b**).